\documentclass{iau} 
\usepackage{graphicx}

\title[IAU 316.~~NGC\,346: Looking in the Cradle of a Massive Star Cluster] 
{NGC\,346: Looking in the Cradle of\\ a Massive Star Cluster}

\author[Dimitrios A. Gouliermis \& Sacha Hony]   
{Dimitrios A. Gouliermis
 \and Sacha Hony}

\affiliation{University of Heidelberg, Centre for Astronomy, Institute for Theoretical Astrophysics, Albert-Ueberle-Str.\,2, 69120 Heidelberg, Germany \\ email: {\tt gouliermis@uni-heidelberg.de}, {\tt sacha.hony@free.fr}}

\pubyear{2015}
\volume{316}  
\jname{Formation, evolution, and survival of massive star clusters}
\editors{C. Charbonnel  \& A. Nota, eds.}
\begin{document}

\maketitle

\begin{abstract}
How does a star cluster of more than few 10,000 solar masses form? We present the case of the cluster NGC\,346 in the Small 
Magellanic Cloud, still embedded in its natal star-forming region N66, and we propose a scenario for its formation, based on observations of 
the rich stellar populations in the region. Young massive clusters  host a high fraction of early-type stars, indicating an extremely high star formation efficiency. 
The Milky Way galaxy hosts several young massive clusters that fill the gap between 
young low-mass open clusters and old massive globular clusters. Only a handful, though, are young enough to study their formation.  
Moreover, the investigation  of their gaseous natal environments suffers from contamination by the Galactic disk. Young massive clusters are very abundant in distant starburst and interacting galaxies, but the 
distance of their hosting galaxies do not also allow a detailed analysis of their formation. The Magellanic Clouds, on the other hand, host young massive clusters in a wide 
range of ages with the youngest being still embedded in their giant HII regions. Hubble Space Telescope imaging of such star-forming 
complexes provide a stellar sampling with a high dynamic range in stellar masses, allowing the detailed study of star formation at scales typical for 
molecular clouds.  Our cluster analysis on the distribution of newly-born stars in N66 shows that star formation in the region proceeds in a clumpy hierarchical fashion, leading to the formation of both a dominant young massive cluster, hosting about half of the observed pre--main-sequence population, and a self-similar dispersed distribution of the 
remaining stars.  We investigate the correlation between stellar surface density (and star formation rate derived from star-counts) and molecular gas surface density (derived 
from dust column density) in order to unravel the physical conditions that gave birth to NGC\,346. 
A power law fit to the data yields a steep correlation between these two parameters with a considerable scatter. The fraction of stellar over the total (gas plus young stars) mass is found to be 
systematically higher within the central 15\,pc (where the young massive cluster is located) than outside, which suggests variations in the star formation 
efficiency within the same star-forming complex. This trend possibly reflects a change of star formation efficiency in N66 between clustered and non-clustered star formation.
Our findings suggest that the formation of NGC\,346 is the combined result of star formation regulated by turbulence and of early
dynamical evolution induced by the gravitational potential of the dense interstellar medium.
\end{abstract}

\firstsection 

\section{Introduction}

It is generally accepted that most stars form in a clustered mode, i.e., in gravitationally bound concentrations. 
Star-forming clusters vary in size, mass, and stellar content, from small compact groups of protostars, still embedded in their natal 
star-forming regions  (\cite[Lada \& Lada 2003]{ladalada03}), to large Young Massive Clusters (YMCs) that cover more than about $10^4$\,M$_\odot$
(\cite[Portegies Zwart et al. 2010]{portegieszwart2010}). The latter, being fundamental contributors to the stellar mass budget of galaxies,   
are valuable laboratories for the investigation of clustered star formation and subsequent evolution. Star clusters themselves are generally not 
formed in isolation. They are the densest assemblings of larger stellar aggregates and complexes \cite[(e.g., Elmegreen 2011)]{elmegreen2011}, 
within a hierarchy of stellar structures, which extends up to galactic scales (e.g., \cite[Gouliermis et al. 2010, 2015]{gouliermis10, gouliermis15}). 

There is only a handful of YMCs close enough to be resolved into their stellar populations. The study of those located in the Milky Way is limited by 
the line-of-site contamination by the Galactic disk, while in the Magellanic Clouds deep multi-band photometric surveys allow the study of star formation 
across the full extent of YMCs natal environments \cite[(e.g., Hony et al. 2010)]{hony2010}. In this study we explore the formation of 
the most massive young stellar cluster in the Small Magellanic Cloud, NGC\,346, located in the star-forming complex LHA\,115-N66 (in short 
N66), the brightest HII region in this galaxy. We exploit the exceptionally rich sample of young stars collected with the Hubble Space Telescope 
(HST) in N66 to address the formation of a YMC from two aspects: (1) The clustering of stars across the length-scale of a giant molecular cloud 
where the YMC forms (\cite[Gouliermis et al. 2014]{gouliermis14}), and (2) the relation between the star formation rate and the environmental 
conditions of the star-forming complex (\cite[Hony et al. 2015]{hony15}).

\begin{figure}[t]
\begin{center}
 \includegraphics[width=3.4in]{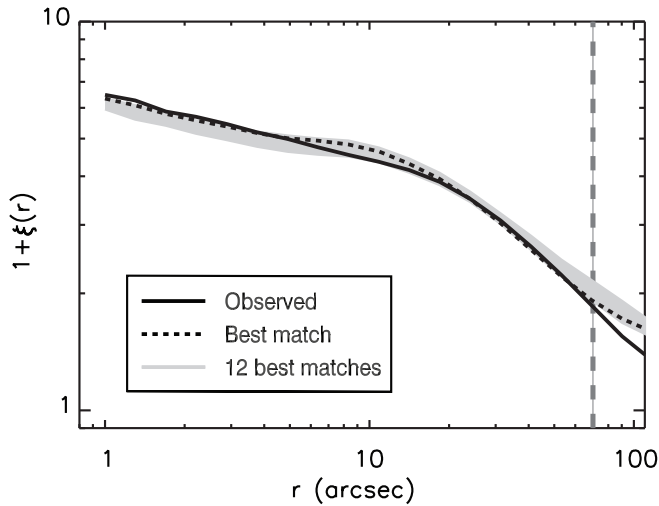} 
 \caption{The observed  ACF of the young stellar population in N66 (solid black line), and the set of 
twelve modeled ACFs (light-grey area), which correspond to the best-representative 
simulated mixed distributions that assume a central compact stellar component and an extended fractal 
one. The ACF of the best-matching model is plotted with a short dashed line. The vertical dashed line corresponds 
to the scale beyond which edge-effects hamper the analysis. Plot from 
\cite[Gouliermis et al. (2014)]{gouliermis14}.}
  \label{fig1}
\end{center}
\end{figure}

\section{The Clustering of Young Stars across N66}

The young population in NGC\,346 consists  of low-mass pre--main-sequence stars, identified from their positions in 
the color-magnitude diagram  (\cite[e.g., Gouliermis 2012] {gouliermis12}), and high-mass upper--main-sequence 
stars (with $m_{\rm 555}-m_{\rm 814} \leq$\,0.0\,mag; 12\,$\lesssim m_{\rm 555} \lesssim$\,17 mag), 
compiling a total sample of 5,150 stars. An age of $\sim$\,5\,Myr has been established for these stars by 
\cite[Mokiem et al. (2006)]{mokiem06}. We quantify the degree of clustering of the stars by using the \textit{autocorrelation function} 
(ACF), based on \cite[Peebles (1980)]{peebles80}. For a two-dimensional self-similar 
distribution this function, $1+\xi$, has a power-law dependency to separation $r$ of the form $1+\xi (r) \propto r^{\eta}$, with the 
exponent $\eta$ being related to the 2D fractal dimension as $D_{2} = \eta +2$. The ACF of the young 
stellar populations of N66 does not follow a single power-law but it has two distinct parts (Fig.\,\ref{fig1}), and therefore 
the stellar distribution is not entirely hierarchical. 

The broken power-law suggests a complex distribution of stars in N66, which is influenced by the 
star cluster NGC\,346. In order to interpret this behavior we applied numerical simulations of centrally-concentrated stellar clusters, following 
an \cite[Elson, Fall \& Freeman (1987)]{eff87} surface density profile, and of three-dimensional fractal stellar distributions, constructed using a 
box-counting technique. From the comparison of the ACFs of synthetic distributions and their combinations we find that the ACF of N66 is best 
reproduced as the composite of two distinct spatial distributions; a centrally-condensed cluster with a core radius of $\sim$\,2.5\,pc and a profile 
index $\gamma \simeq$\,2.27, and a self-similar stellar distribution with a 3D fractal dimension $D_3\approx$\,2.3  (Fig.\,\ref{fig1}). About 40\% 
of the total young stellar population belongs to the cluster, while the remaining is spread across the whole extent of the observed star-forming 
complex. The separation of $\sim$\,6\,pc where the ACF power-law breaks indicates the length-scale where the clustering  of stars 
changes from one pattern to the other.

Our findings confirm the appearance of substructure in the region, established in a previous study of ours \cite[(Schmeja et al. 2009)]{schmeja09}, 
as the result of hierarchical stellar distribution. The fractal dimension $D_3 \simeq 2.3$, derived for the self-similar stellar component, fits very well 
to the value established from numerical experiments of supersonic isothermal turbulence (\cite[Federrath et al. 2009]{federrath09}), implying that the 
observed hierarchy is inherited from the turbulent interstellar gas. On the other hand, numerical simulations find a higher 
degree of clumping for sink particles formed from turbulent molecular clouds ($D_3 \sim$\,1.6; \cite[Girichidis et al. 2012]{girichidis12}). 
The differences in the derived fractal dimensions may reflect discrepancies between the methods used for their measurement, 
but they may as well demonstrate that {\em stars have a different spatial distribution than the gas from which they formed}. Even if stars were
originally distributed according to a turbulent-driven hierarchy, N-body simulations  suggest that a moderate amount of dynamical interactions will partly 
erase substructure, preventing a star-forming region from retaining a strong signature of the primordial ISM distribution (\cite[e.g., Goodwin \& Whitworth 2004, 
Parker et al. 2014]{goodwin04, parker14}). The substructure in N66 is unlikely to have formed after the onset of star formation, and it is therefore  
{\em the observed fractal dimension is an upper limit to the primordial value}. Dynamics may have facilitated the formation of the 
YMC within the original self-similar distribution of stars. Kinematic studies in the region will certainly shed light to the importance of dynamical 
evolution in the formation of NGC\,346.

In conclusion, our findings suggest that {\em star formation in N66 takes place in a clustered, as well as a dispersed mode}, revealed by the bimodal -- centrally-condensed and self-similar -- 
clustering of newly-born stars in the region \cite[(Gouliermis et al. 2014)]{gouliermis14}.


\section{The Relation between Gas and Young Stars in N66}

To understand the formation of NGC\,346 in relation to the surrounding interstellar medium (ISM), we examine the ISM properties 
as a function of the surface density of young stars across N66 (\cite[Hony et al. 2015]{hony2015}). The young stellar sample used 
for the cluster analysis discussed in the previous section is also used for the derivation of the {\em surface density of the star formation rate} 
($\Sigma_{\rm SFR}$) through star counts. We  employ the {\em ``dust method''} to determine the ISM column density, i.e. using the infrared (IR) to submillimetre (submm) dust continuum emission to derive via radiative transfer modeling of its spectral energy
 distribution (SED) the dust surface density, which we transform into {\em gas surface density} ($\Sigma_{\rm gas}$). This method has the advantage of not being sensitive to the state of the gas (ionized, neutral or molecular) as long as dust and gas are well mixed. This is important because at the age of the young stars in N66 the parental molecular cloud  may have been subject to significant photodissociation and using a molecular gas tracer (such as CO) may lead to an incomplete view of the total gas surface densities.

In our analysis, we use IR-to-submm photometric maps from {\em Spitzer} (\cite[SAGE-SMC; Gordon et al. 2011]{gordon11}) and {\em Herschel} (\cite[HERITAGE; 
Meixner et al. 2013]{meixner13}), which we convolve to the lowest available resolution of 20$^{\prime\prime}$, corresponding to the SPIRE 250\,$\mu$m beam. We 
have also convolved  the stellar density map, which has an exceedingly good intrinsic angular resolution, to this effective spatial resolution. After convolution, all maps 
have been re-projected on the same pixels scheme with pixel size of 20$\times$20\,arcsec$^{2}$ in order to have spatially independent measurements and to allow pixel-by-pixel 
SED extraction and comparison between the stellar densities and derived dust surface densities. We thus study the small-scale correlation between newly formed stars and the gas reservoir in 115 pixels (of $\sim$\,6\,pc size), while covering the entire star-forming complex.  The conversion of dust to gas surface density was made with the use of a 
gas-to-dust ratio ($r_{\rm gd} =$\,1250), calibrated using gas tracers on large scales. The SFR map is derived from the observed number of young stars in each pixel of the stellar density map using a mass per detected star of 4.3\,M$_{\odot}$, derived 
using the stellar Initial Mass Function of the region (\cite[Sabbi et al. 2008]{sabbi08}), and a star formation duration of 5\,Myr \cite[(Mokiem et al. 2006)]{mokiem06}. This SFR from star-counts is found to be more reliable than indirect SFR tracers (e.g., H$\alpha$, 24\,$\mu$m emission) on parsec scales. We show that these tracers break down on small scales ($<$\,100\,pc) by up to a factor of 10.

\begin{figure}[t]
\begin{center}
 \includegraphics[width=3.4in]{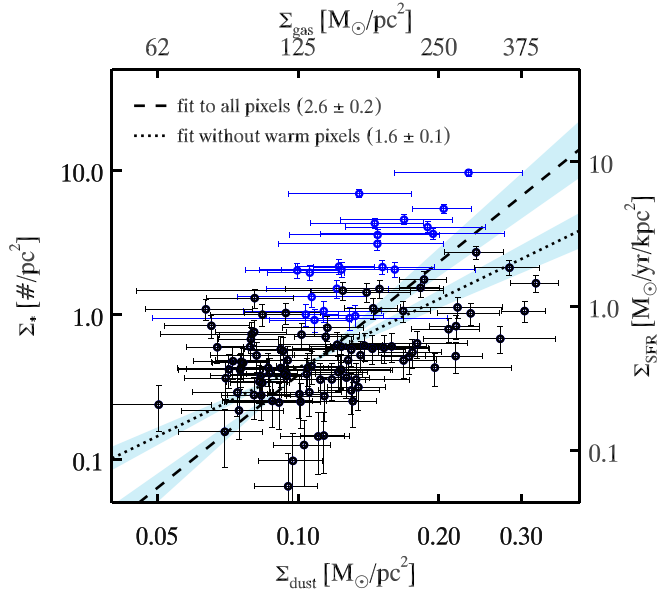} 
 \caption{Scatter plot of surface density of young stars versus surface density of dust (and the corresponding SFR and gas surface density) 
 per pixel of 20$\times$20\,arcsec$^{2}$ (6$\times$6\,pc$^{2}$) in the star-forming complex N66. The blue points have a strong warm dust 
 component in their SED ($\Sigma_{24\,{\mu}{\rm m}}$/$\Sigma_{250\,{\mu}{\rm m}}$ [F$_\nu$/F$_\nu$] $>$ 0.3). The lines-of-sight selected 
 using this color criterion clearly occupy a separate region of the diagram. The dashed and dotted lines are the power-law fits to the data 
 with or without the warm pixels, respectively. The blue areas correspond to the 2$\sigma$ uncertainties of the best-fitting exponents, which are
 given in parentheses. There is a clear correlation between $\Sigma_{\rm SFR}$ and $\Sigma_{\rm gas}$ with some considerable scatter, 
 suggesting a dependence of the SFR to the density of the gas reservoir. Plot from \cite[Hony et al. (2015)]{hony15}.}
   \label{fig2}
\end{center}
\end{figure}

\begin{figure}[t]
\begin{center}
 \includegraphics[width=3.4in]{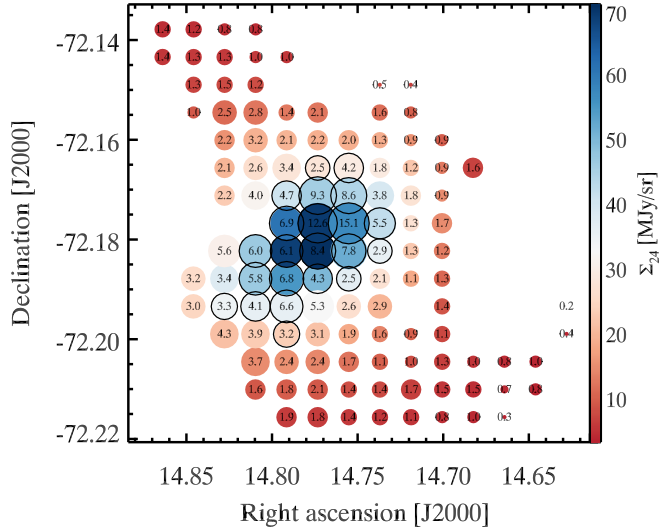} 
 \caption{Map of N66 showing the spatial distribution of the derived stellar mass fraction. 
 The size of each point is dependent on $f_{\star}$ and its color depends on $\Sigma_{24\,{\mu}{\rm m}}$. 
 The mass fraction is indicated in each point. Stellar mass fraction clearly peaks on the stellar cluster where 
 the dust is heated to higher temperatures due to the intense radiation field. The warm pixels identified in
Fig.~\ref{fig2} are encircled. Plot from \cite[Hony et al. (2015)]{hony15}.}
   \label{fig3}
\end{center}
\end{figure}

A correlation between the stellar surface density, $\Sigma_{\star}$, and the dust surface density, $\Sigma_{\rm dust}$, and between the corresponding 
parameters, $\Sigma_{\rm SFR}$ and $\Sigma_{\rm gas}$, is found with considerable scatter (Fig.\,\ref{fig2}). A power-law fit to the data yields a 
steep relation with an exponent of 2.6\,$\pm$\,0.2. We find that sight-lines towards the central $\lesssim$\,15\,pc, where NGC\,346 is located, exhibit 
systematically high values of  $\Sigma_{\star}$/$\Sigma_{\rm dust}$ by a factor of $\sim$5 compared to the rest of the complex \cite[(Hony et al. 2015)]{hony2015}.  
We investigate the variations in terms of the mass fraction of young stars, $f_{\star} = \Sigma_{\star}/\left(\Sigma_{\star}+\Sigma_{\rm gas}\right)$,
across the star-forming complex. The spatial distribution of the observed stellar mass fraction
shows that it is systematically high (with a maximum of 15 per cent) within the central 15\,pc of the complex, where the YMC is located (Fig.\,\ref{fig3}). These very high values 
correspond to very high stellar surface densities, but \emph{not} to low gas column densities.  The stellar mass fraction becomes lower outside the central 15\,pc, 
across the remaining surveyed area (66 per cent of the pixels), where it has values systematically below 2 per cent. There is a clear monotonic decrease in 
$f_{\star}$ with radial distance from the center, which can be approximated by a power law with exponent $-0.7$. These variations show also a clear dependence 
of $f_{\star}$ on other measurable quantities, such as the surface density of hot dust emission at 24\,$\mu$m. These findings indicate clearly a higher SFE in the
inner 15\,pc area, which is dominated by the young stars belonging to the YMC, while the outer parts, characterized by lower SFE, are those dominated
by the dispersed self-similar distribution of young stars (Sect.\,2). 


Our analysis allow us to place these measurements on a Schmidt-Kennicutt (SK) diagram. We find that individual pixels fall systematically above the 
fiducial SK relation for $\Sigma_{\rm SFR}$ versus $\Sigma_{{\rm HI}+{\rm H}_{2}}$ for integrated disk-galaxies \cite[(Kennicutt \& Evans 2012)]{kennicutt12}
 by on average a factor of $\sim$7. 
This behavior is consistent with the results by \cite[Heiderman et al. (2010)]{heiderman10}, who found that the measured SFR lies above the galaxy 
averages when `zooming-in' to parsec-scale Galactic star-forming clumps. This is probably caused by less dense gas, which is inefficient in
forming stars, and which is included in the galaxy-scale averages but not measured when `zooming-in' on individual clumps.
In our analysis, on the other hand, even though we analyze 6$\times$6\,pc$^{2}$ sized regions, we cover the complete star-forming 
complex and find that the entire region (of $\sim$50\,pc), being quite active, lies consistently above the SK relation. For N66 one should `zoom-out' 
beyond 50\,pc to probe this less dense, non-star-forming gas. Indeed, averaging over a larger area (90\,pc in radius) we derive a measurement 
for N66 that lies closer to the SK-relation, but which still remains high by a factor of $\sim$3.

In conclusion, we find an {\em above-average star formation activity in N66}, and that the observed correlations between stellar and ISM properties reflect {\em a change in star formation efficiency between clustered and non-clustered star-formation} 
within the same star-forming complex \cite[(Hony et al. 2015)]{hony2015}.


\acknowledgements
Based on observations made with NASA/ESA Hubble Space Telescope, 
obtained from the data archive at the Space Telescope Science Institute (STScI), 
Spitzer Space Telescope, {\em Herschel} and Atacama Pathfinder Experiment 
(APEX). STScI is operated by the Association of Universities for Research in 
Astronomy, Inc. under NASA contract NAS\,5-26555. {\em Herschel} is an ESA 
space observatory with science instruments provided by European-led Principal 
Investigator consortia and with important participation from NASA. Spitzer Space 
Telescope is operated by the Jet Propulsion Laboratory, California Institute of 
Technology under a contract with NASA. APEX  is a collaboration between the
Max-Planck-Institut f\"{u}r Radioastronomie, the European Southern Observatory, 
and the Onsala Space Observatory. The authors kindly acknowledge support from the 
{\em German Research Foundation} (DFG) through the individual grants GO 1659/3-1 
and  GO 1659/3-2, 
and the collaborative research project SFB881 ``The Milky Way System'' (subprojects B1, 
B2, and B5) respectively.



\begin{thebibliography}{}



\bibitem[Elmegreen(2011)]{elmegreen2011} 
Elmegreen B. G., 2011, in: C. Charbonnel \& T. Montmerle (eds.), \textit{Star Formation in the Local Universe}, 
EAS Publications Series Vol. 51 (Cambridge: Cambridge Univ. Press), p.\,31


\bibitem[Elson, Fall \& Freeman (1987)]{eff87} 
Elson, R.~A.~W., Fall, S.~M., \& Freeman, K.~C.\ 1987, \textit{ApJ}, 323, 54 

\bibitem[Federrath et al.(2009)]{federrath09} 
Federrath, C., Klessen, R.~S., \& Schmidt, W.\ 2009, \textit{ApJ}, 692, 364 

\bibitem[Girichidis et al.(2012)]{girichidis12} 
Girichidis, P., Federrath, C., Allison, R., et al.\ 2012, \textit{MNRAS}, 420, 3264 

\bibitem[Goodwin \& Whitworth(2004)]{goodwin04} 
Goodwin, S.~P., \& Whitworth, A.~P.\ 2004,  \textit{A\&A}, 413, 929

\bibitem[Gordon et al.(2011)]{gordon11} 
Gordon, K.~D., Meixner, M., Meade, M.~R., et al.\ 2011, \textit{AJ}, 142, 102 

\bibitem[Gouliermis(2012)]{gouliermis12} 
Gouliermis, D.~A.\ 2012, \textit{Space Sci. Revs}, 169, 1 

\bibitem[Gouliermis et al.(2006)]{gouliermis06}
Gouliermis, D. A., Dolphin A. E., Brandner W., Henning T., 2006, \textit{ApJS}, 166, 549

\bibitem[Gouliermis et al.(2010)]{gouliermis10} Gouliermis, D.~A., 
Schmeja, S., Klessen, R.~S., et al.\ 2010, \textit{ApJ}, 725, 1717

\bibitem[Gouliermis et al.(2014)]{gouliermis14} Gouliermis, D.~A., 
Hony, S., \& Klessen, R.~S.\ 2014, \textit{MNRAS}, 439, 3775

\bibitem[Gouliermis et al.(2015)]{gouliermis15} Gouliermis, D.~A., 
Thilker, D., Elmegreen, B.~G., et al.\ 2015, \textit{MNRAS}, 452, 3508 

\bibitem[Heiderman et al.(2010)]{heiderman10} 
Heiderman, A., Evans, N.~J., II, Allen, L.~E., Huard, T., \& Heyer, M.\ 2010,  \textit{ApJ}, 723, 1019 

\bibitem[Hony et al.(2010)]{hony2010} 
Hony, S., Galliano, F., Madden, S.~C., et al.\ 2010, \textit{A\&A}, 518, L76 

\bibitem[Hony et al.(2015)]{hony2015} Hony, S., Gouliermis, 
D.~A., Galliano, F., et al.\ 2015, \textit{MNRAS}, 448, 1847

\bibitem[Kennicutt \& Evans(2012)]{kennicutt12} 
Kennicutt, R.~C., \& Evans, N.~J.\ 2012, \textit{ARAA}, 50, 531 


\bibitem[Lada \& Lada(2003)]{ladalada03} 
Lada, C.~J., \& Lada, E.~A.\ 2003, \textit{ARAA}, 41, 57 

\bibitem[Meixner et al.(2013)]{2013AJ....146...62M} Meixner, M., Panuzzo, 
P., Roman-Duval, J., et al.\ 2013, \textit{AJ}, 146, 62 

\bibitem[Mokiem et al.(2006)]{mokiem06} 
Mokiem, M.~R., de Koter, A., Evans, C.~J., et al.\ 2006, \textit{A\&A}, 456, 1131 

\bibitem[Parker et al.(2014)]{parker14} Parker, R.~J., Wright, 
N.~J., Goodwin, S.~P., \& Meyer, M.~R.\ 2014, \textit{MNRAS}, 438, 620

\bibitem[Peebles(1980)]{peebles80}
{Peebles}, P.~J.~E. 1980, \textit{The large-scale structure of the universe} (Princeton Univ. Press)

\bibitem[Portegies Zwart et al.(2010)]{portegieszwart2010} 
Portegies Zwart, S.~F., McMillan, S.~L.~W., \& Gieles, M.\ 2010, \textit{ARAA}, 48, 431

\bibitem[Sabbi et al.(2008)]{sabbi08} 
Sabbi, E., Sirianni, M., Nota, A., et al.\ 2008, \textit{AJ}, 135, 173 

\bibitem[Schmeja et al.(2009)]{schmeja09} 
Schmeja, S., Gouliermis, D.~A., \& Klessen, R.~S.\ 2009, \textit{ApJ}, 694, 367 


\end{thebibliography}
\end{document}